\documentclass{ifacconf}

\usepackage{graphicx}      
\usepackage{natbib}        
\usepackage{amsmath}
\usepackage{amssymb}
\usepackage{stackengine}
\usepackage{url}

\DeclareMathOperator{\sinc}{sinc}

\newtheorem{mydef}{Definition}
\newtheorem{mythm}{Theorem}
\newtheorem{assumption}{Assumption}

\newtheorem{myproof}[mydef]{Proof}

\newcommand{\qedsymbol}{\rule{0.7em}{0.7em}}

\begin{document}
\begin{frontmatter}

\title{Linear Parameter Varying Representation of a class of MIMO Nonlinear Systems} 

\thanks[footnoteinfo]{Maarten Schoukens is supported by the European Union's Horizon 2020 research and innovation programme under the Marie Sklodowska-Curie Fellowship (grant agreement nr. 798627). Furthermore, R. T\'oth is supported by the European Research Council (ERC) under the European Union's Horizon 2020 research and innovation programme (grant agreement nr. 714663).}

\author[First]{Maarten Schoukens} 
\author[First]{Roland T\'oth}

\address[First]{Control Systems, Eindhoven University of Technology, Eindhoven, The Netherlands (e-mail: m.schoukens@tue.nl, r.toth@tue.nl).}

\begin{abstract}                
	Linear parameter-varying (LPV) models form a powerful model class to analyze and control a (nonlinear) system of interest. Identifying an LPV model of a nonlinear system can be challenging due to the difficulty of selecting the scheduling variable(s) a priori, especially if a first principles based understanding of the system is unavailable. Converting a nonlinear model to an LPV form is also non-trivial and requires systematic methods to automate the process. 
	Inspired by these challenges, a systematic LPV embedding approach starting from multiple-input multiple-output (MIMO) linear fractional representations with a nonlinear feedback block (NLFR) is proposed. This NLFR model class is embedded into the LPV model class by an automated factorization of the (possibly MIMO) static nonlinear block present in the model. As a result of the factorization, an LPV-LFR or an LPV state-space model with affine dependency on the scheduling is obtained. This approach facilitates the selection of the scheduling variable and the connected mapping of system variables. Such a conversion method enables to use nonlinear identification tools to estimate LPV models.
	The potential of the proposed approach is illustrated on a 2-DOF nonlinear mass-spring-damper example.
\end{abstract}


\end{frontmatter}

\section{Introduction}
	The Linear-Parameter Varying (LPV) framework offers a powerful tool set to model and control nonlinear systems \citep{Mohammadpour2012}. Many control applications depend on the availability of high-quality LPV models, fueling the need for LPV identification algorithms. LPV identification has been studied in detail \citep{Toth2010,Santos2011} in the past. However, in the majority of these works it is assumed that the scheduling signals are known a priori, relying on the user's expertise to design them for the considered system. Furthermore, often the noise in the measured version of the scheduling signals are left unattended in LPV identification, leading to a possible bias of the estimates (for approaches to handle this see \citep{Piga2015}).
		
	Embedding nonlinear models into the LPV model class offers an alternative approach to obtain LPV models of nonlinear systems without having to identify an LPV model directly. This avoids the selection of appropriate scheduling signals during LPV identification: the scheduling signal(s) are obtained as a result of the nonlinear embedding. Although embedding of nonlinear systems into the LPV framework is a popular approach in nonlinear systems control, only a few systematic embedding methods are discussed in the literature \citep{Chisci2003,Toth2010,Young2011,Mohammadpour2012,Abbas2014,Abbas2017}. 
	
\begin{figure}[bt]
		\centering
			\includegraphics[width=0.95\columnwidth]{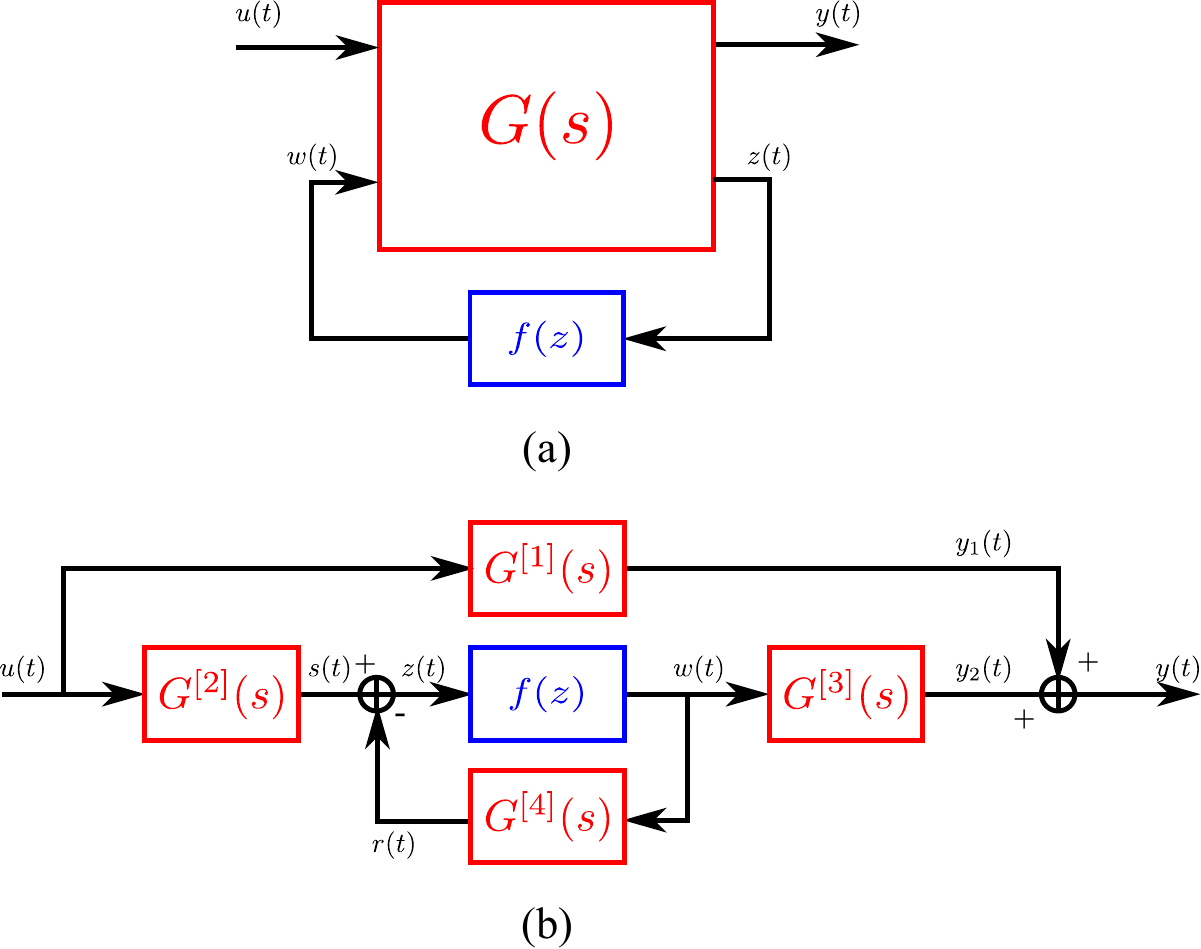}
		\caption{The NLFR structure represented by (a) a MIMO LTI block $G$, and (b) an equivalent block-oriented structure with 4 MIMO LTI blocks $G^{[i]}$ and the MIMO static nonlinear function $f(z)$.}
		\label{fig:LFR}
	\end{figure}

	LPV embedding of NLFR structures, sometimes called Lur'e systems, has been studied in prior works \citep{Seron2015,Hanafi2018,SchoukensM2018}. However, these works are limited to single-input-single-output static nonlinearities. This paper presents a systematic embedding approach for MIMO nonlinear systems represented by NLFRs with a MIMO static nonlinear block (see Figure~\ref{fig:LFR}) \citep{SchoukensM2017b}. It is shown how one can embed this model in an automated and systematic way into a MIMO LPV representation such that an LPV-LFR, or alternatively an affine state-space LPV model results, without introducing new singularity points in the representation.
	
	The considered MIMO NLFR system class is discussed in Section~\ref{sec:LFRClass}. Section~\ref{sec:Embedding} discusses the embedding of these nonlinear systems into an LPV representation using a systematic embedding algorithm. Finally, a 2-DOF nonlinear mass-spring-damper system is analyzed to illustrate the developed embedding process.
	
\section{NLFR System Class} \label{sec:LFRClass}
	In this work, we considered the class of nonlinear systems that can be represented by a MIMO continuous-time NLFR. The NLFR system class is a general system class comprising a, possibly MIMO, localized static nonlinearity (see Figure~\ref{fig:LFR}). Identification of such NLFR structures is considered in \citep{Vandersteen1999,Hsu2008,Novara2011,Vanbeylen2013}. Generality of this representation is based upon the fact that widely used block-oriented Hammerstein, Wiener and Wiener-Hammerstein structures are special cases of the NLFR structure (see \citep{Giri2010,SchoukensM2017b} for more information on the identification of block-oriented structures and some of their applications).
	
	The input-output relation of a continuous-time MIMO NLFR can be expressed via a (minimal) state-space form as:
	\begin{align}
		\begin{split} \label{eq:LFRSS1}
		\dot{x}(t) &= A x(t) + \begin{bmatrix}B_\mathrm{w} & B_\mathrm{u}\end{bmatrix} \begin{bmatrix} w(t) \\ u(t) \end{bmatrix}, \\
		\begin{bmatrix}z(t) \\ y(t)\end{bmatrix} &= \begin{bmatrix}C_\mathrm{z} \\ C_\mathrm{y}\end{bmatrix} x(t) + \begin{bmatrix}D_\mathrm{zw} & D_\mathrm{zu} \\ D_\mathrm{yw} & D_\mathrm{yu} \end{bmatrix} \begin{bmatrix} w(t) \\ u(t) \end{bmatrix}, \\
		w(t) &= f(z(t)),
		\end{split}
	\end{align}
	where $u(t) \in \mathbb{R}^{n_u \times 1}$ is the input of the NLFR, $w(t) \in \mathbb{R}^{n_w \times 1}$ is the output of the static nonlinearity, $x(t) \in \mathbb{R}^{n_x \times 1}$ are the system states, $y(t) \in \mathbb{R}^{n_y \times 1}$ is the output of the NLFR, $z(t) \in \mathbb{R}^{n_z \times 1}$ is the input of the static nonlinearity, $f(z(t))$, $\mathbb{R}^{n_z} \rightarrow \mathbb{R}^{n_w}$ is a multi-dimensional, static and bounded nonlinear function, and $A$, $\ldots$ $D_\mathrm{yu}$ are real constant matrices of appropriate dimensions. 
	
	The signal transfers corresponding to eq.~\eqref{eq:LFRSS1} can be also equivalently represented by MIMO LTI transfer matrices $G^{[i]}$:
	\begin{align} \label{eq:LTITF}
	\begin{split}
		G^{[1]}(s) &= C_\mathrm{y}(sI-A)^{-1}B_\mathrm{u} + D_\mathrm{yu},\\
		G^{[2]}(s) &= C_\mathrm{z}(sI-A)^{-1}B_\mathrm{u} + D_\mathrm{zu},\\
		G^{[3]}(s) &= C_\mathrm{y}(sI-A)^{-1}B_\mathrm{w} + D_\mathrm{yw},\\
		G^{[4]}(s) &= C_\mathrm{z}(sI-A)^{-1}B_\mathrm{w} + D_\mathrm{zw},
	\end{split}
	\end{align}
	where $s \in \mathbb{C}$ denotes the complex frequency (Laplace variable).
	
	Substituting $z(t)$ into \eqref{eq:LFRSS1} results in:
	\begin{align}
		\begin{split} \label{eq:LFRSS2}
		\dot{x}(t) &\!=\! A x(t) \!+\! B_\mathrm{u} u(t) \\
			   &\quad +\! B_\mathrm{w} f(C_\mathrm{z} x(t) \!+\! D_\mathrm{zu} u(t) \!+\! D_\mathrm{zw} w(t)), \\
		y(t)   &\!=\! C_\mathrm{y} x(t) \!+\! D_\mathrm{yu} u(t) \\
			   &\quad +\! D_\mathrm{yw} f(C_\mathrm{z} x(t) \!+\! D_\mathrm{zu} u(t) \!+\! D_\mathrm{zw} w(t)), \\
		w(t) &\!=\! f(C_\mathrm{z} x(t) + D_\mathrm{zu} u(t) + D_\mathrm{zw} w(t)).
		\end{split}
	\end{align}
	
	To simplify our problem setting the following assumption is taken:
	\begin{assumption} \label{ass:proper}
		$G^{[4]}$ is strictly proper (in other words, $D_\mathrm{zw}=0$).
	\end{assumption}
	Under Assumption~\ref{ass:proper}, it is possible to eliminate $w(t)$. This results in the following simplified expression:
	\begin{align}
		\begin{split} \label{eq:LFRSS3}
		\dot{x}(t) &\!=\! A x(t) \!+\! B_\mathrm{u} u(t) \!+\! B_\mathrm{w} f(C_\mathrm{z} x(t) \!+\! D_\mathrm{zu} u(t)), \\
		y(t)   &\!=\! C_\mathrm{y} x(t) \!+\! D_\mathrm{yu} u(t) +\! D_\mathrm{yw} f(C_\mathrm{z} x(t) \!+\! D_\mathrm{zu} u(t)).
		\end{split}
	\end{align}
	Analogously, similar expressions can be obtained in discrete-time form in a straightforward manner.
	
	Assumption~\ref{ass:proper} forces the nonlinearity $f(z)$ to be explicit. The presence of a direct feedthrough term $D_\mathrm{zw}$ would allow the nonlinearity to present itself in as an implicit function.
	
\section{LPV Embedding} \label{sec:Embedding}
	\subsection{Embedding Concept}
	The static nonlinearity that is present in the nonlinear model needs to be factorized to obtain an LPV representation. Multiple factorization approaches are possible. Here, the nonlinear function $f(z(t))$ is decomposed as $\bar{f}(z(t))z(t) + c$.
	
	Denote $\mathbb{P}$ as the set of values that can be reached by $z(t)$ for a specified class of inputs $u(t) \in \mathbb{U}$ with $t \in [0 \: \infty)$ and initial conditions $x(0) \in \mathbb{X}$.	
	\begin{assumption} \label{ass:SNL} The static nonlinear function $f(z(t))$ can be represented as: $\bar{f}(z(t))z(t) + c$, such that $\bar{f}(z(t))$ does not contain singular points for $z(t) \in \mathbb{P}$, and $c$ is finite. \end{assumption}
	
	This assumption excludes for instance functions $f(z(t))$ that have singularities in the region of interest (e.g. $\frac{1}{z}$, if $z=0$ lies within the range of interest). Note that decomposition of a function in the form given in Assumption~\ref{ass:SNL} is non-unique. The function $\bar{f}(z(t))$, $\mathbb{R}^{n_z} \rightarrow \mathbb{R}^{n_w \times n_z}$ is called the scheduling map in the sequel.
	
	\subsection{Constant Offset}
	
	\begin{assumption} \label{ass:stability}
		All linear subsystems $G^{[i]}$ (see Eq.~\eqref{eq:LTITF}) are bounded-input bounded-output stable. 
	\end{assumption}
	Assumption~\ref{ass:stability} is only required for the offset propagation algorithm. If no constant offset ($f(0) = 0$) is present in the static nonlinearity, this assumption is not required. Note that this assumption does not impose stability of the total NLFR system.
	
	Under Assumptions~\ref{ass:SNL} and \ref{ass:stability}, the NLFR structure with $D_\mathrm{zw}=0$ can be represented by:
	\begin{align}
		\begin{split}
		\dot{x}(t) &= A x(t) + B_\mathrm{u} \tilde{u}(t) + B_\mathrm{w} \tilde{f}(C_\mathrm{z} x(t) + D_\mathrm{zu} \tilde{u}(t)), \\
		\tilde{y}(t)   &= C_\mathrm{y} x(t) + D_\mathrm{yu} \tilde{u}(t) + D_{3} \tilde{f}(C_\mathrm{z} x(t) + D_\mathrm{zu} \tilde{u}(t)),
		\end{split}
	\end{align}
	where:
	\begin{align} \label{eq:offsetLPV}
		\begin{split}
		\tilde{f}(z(t)) &= \bar{f}(z(t))z(t),  \\
		\tilde{u}(t) &= u(t) - {G_0^{[2]}}^{-1}{G_0^{[4]}} c,  \\
		\tilde{y}(t) &= y(t) - \left( G_0^{[3]} + G_0^{[1]}{G_0^{[2]}}^{-1}{G_0^{[4]}} \right) c,
		\end{split}
	\end{align}
	where $G^{[i]}_0$ is obtained by evaluating the steady-state gain of $G^{[i]}(s)$, i.e. $G^{[i]}_0 = \lim_{s \rightarrow 0} G^{[i]}(s)$.
	Eq.~\eqref{eq:offsetLPV} to hold, all ${G_0^{[2]}}^{-1}{G_0^{[4]}}$ and $\left( G_0^{[3]} + G_0^{[1]}{G_0^{[2]}}^{-1}{G_0^{[4]}} \right)$ should exist and be finite. In the case $n_z \leq n_u$ and the rank of $G_0^{[2]}$ is equal to $n_z$, the Moore-Penrose left-inverse of the matrix $G_0^{[2]}$ can be used \citep{Golub1996}. This is however a conservative requirement. It suffices that the following assumption holds:
	\begin{assumption} \label{ass:offsetmitigation}
		The vector $G_0^{[4]}c$ is part of the column space of the matrix $G_0^{[2]}$. 
	\end{assumption}
	Under Assumption~\ref{ass:offsetmitigation} one can always find a vector $d$ such that $G_0^{[2]}d = G_0^{[4]}c$, resulting in the following offset mitigation equations:
	\begin{align} \label{eq:offsetLPV}
		\begin{split}
		\tilde{u}(t) &= u(t) - d,  \\
		\tilde{y}(t) &= y(t) - G_0^{[3]}c - G_0^{[1]}d,
		\end{split}
	\end{align}
	Note that Assumption~\ref{ass:offsetmitigation} is only required for the offset propagation algorithm.
	
	The constant offset at both the input and the output can be dealt with during the LPV control design process as a disturbance or by using input or output trimming methods.
	
	\subsection{Factorization}
	The time-dependent notation $z(t)$ is simplified to $z$ to lighten the notation in this section.
	
	This subsection introduces a systematic approach to perform the factorization $\tilde{f}(z) = \bar{f}(z)z$ for $z \in \mathbb{P}$. One can write: 
	\begin{align}
		\begin{split}
		\tilde{f}(z) &= \tilde{f}(z_1,z_2,\ldots,z_{n_z}) \\
					 &= \sum_{i=1}^{n_z} \bar{f}_{i}(z_1,z_2,\ldots,z_{i})z_{i}.
		\end{split}
	\end{align}
	
	\begin{assumption} \label{ass:Derivative} All the first order partial derivatives:
	\begin{align}
		\left. \frac{\partial \tilde{f}(z)}{\partial z_i}\right|_{z=(z_1,\ldots,z_{i\!-\!1},0,\ldots,0)}
	\end{align} 
	of $\tilde{f}(z)$, $z \in \mathbb{P}$ exist.
	\end{assumption}
	This assumption holds, for example, for all continuously differentiable functions $\mathcal{C}_1^{n_w} (\mathbb{R}^{n_z})$.
	
	Under assumptions \ref{ass:SNL} and \ref{ass:Derivative}, the functions $\bar{f}_{i}$ are given by:
	\begin{align} \label{eq:fi}
	\begin{split}
		&\bar{f}_{i}(z_1,z_2,\ldots,z_{i}) = \\
		&\begin{cases} \frac{\tilde{f}(z_1,\ldots,z_{i},0,\ldots,0) \!-\! \tilde{f}(z_1,\ldots,z_{i-1},0,\ldots,0)}{{z_i}} \\ \qquad \qquad \qquad \qquad \qquad \qquad \qquad \qquad \quad \textrm{if} \; z_{i} \!\neq\! 0 \\
	    \left.\frac{\partial \tilde{f}(z_1,\ldots,z_{i},0,\ldots,0)}{\partial z_i}\right|_{z=(z_1,\ldots,z_{i\!-\!1},0,\ldots,0)} \!\!\! \textrm{if} \; z_{i} \!=\! 0 \end{cases}
	\end{split}
	\end{align}
	and
	\begin{align}
		\bar{f}_{1}(z_1) = \begin{cases} \frac{\tilde{f}(z_1,0,\ldots,0)}{{z_1}} &\quad \textrm{if} \; z_{1} \neq 0 \\
		\left.\frac{\partial \tilde{f}(z_1,0,\ldots,0)}{\partial z_1}\right|_{z=0}  &\quad \textrm{if} \; z_{1} = 0 \end{cases} 
	\end{align}
	
	\begin{mythm} 
		The function $\bar{f}_i(z_1,\ldots,z_i)$ is continuous at all $z\in \mathbb{P}$ if $\tilde{f}(z)$ is continuous at all $z = (z_1,\ldots,z_{i-1},0,\ldots,0)$, $z\in \mathbb{P}$ and if Assumption \ref{ass:Derivative} holds.
	\end{mythm}
	\begin{myproof}
		The quotient of two continuous functions $\frac{h_1(z)}{h_2(z)}$ is continuous everywhere, except in the zeros of $h_2(z)$.
		
		The function $\bar{f}_i(z_1,\ldots,z_i)$ is given by Eq.~\eqref{eq:fi}, where both the numerator and the denominator are formed by continuous functions by assumption. Hence, $\bar{f}_i(z_1,\ldots,z_i)$ is continuous everywhere, except in the point $z_i = 0$.
		
		For  $\bar{f}_i(z_1,\ldots,z_i)$ to be continuous in $z_i=0$ we need that:
		\begin{align} \label{eq:lim}
			\lim_{z_i \rightarrow 0}\bar{f}_i(z_1,\ldots,z_i) = \bar{f}_i(z_1,\ldots,z_{i-1},0)
		\end{align}
		and Eq.~\eqref{eq:lim} should be finite. It is easy to observe that:
		\begin{align} \label{eq:lim}
		\begin{split}
			&\lim_{z_i \rightarrow 0}\bar{f}_i(z_1,\ldots,z_i) \\
			&= \left.\frac{\partial \tilde{f}(z_1,\ldots,z_{i},0,\ldots,0)}{\partial z_i}\right|_{z=(z_1,\ldots,z_{i-1},0,\ldots,0)}
		\end{split}
		\end{align}
		which exists and is finite (see Assumption \ref{ass:Derivative}).
		
		Hence, $\bar{f}_i(z_1,\ldots,z_i)$ is continuous everywhere and no singularities are introduced by the factorization.  
		
		\raggedleft{\qedsymbol}
	\end{myproof}
		
	It can be observed that for a polynomial (or rational) function $\tilde{f}(z)$, also the functions $\bar{f}_{i}(z_1,z_2,\ldots,z_{i})$ are polynomial (rational) (remember that $\tilde{f}(z)$ never contains a constant term: $\tilde{f}(0) = 0$).
	
	In case the function $\tilde{f}(z)$ is not partially differentiable in the points of interest, but its left and right derivative are finite in those points, the partial derivatives can be replaced by a finite constant without introducing a singularity in $\bar{f}(z)$. However, the continuity of $\bar{f}(z)$ is lost.
	
	The proposed factorization is not unique. By changing the order in which the variables $z_i$ are considered, different factorizations of the multivariate nonlinearity could be obtained.
	
	\subsection{Embedded Representation}
	Under Assumption~\ref{ass:SNL} we can represent the NLFR structure as an LPV affine state-space representation with an additional constant offset at the input and/or output:
	\begin{align} \label{eq:LFRSS4}
		\begin{split}
		\!\!\!\!\dot{x}(t) \!&=\! A x(t) \!+\! B_\mathrm{u} \tilde{u}(t) \!+\! B_\mathrm{w} p(t)\left(C_\mathrm{z} x(t) \!+\! D_\mathrm{zu} \tilde{u}(t)\right), \\
		\!\!\!\!\tilde{y}(t)   \!&=\! C_\mathrm{y} x(t) \!+\! D_\mathrm{yu} \tilde{u}(t) \!+\! D_\mathrm{yw} p(t)\left(C_\mathrm{z} x(t) \!+\! D_\mathrm{zu} \tilde{u}(t)\right),
		\end{split}
	\end{align} 
	where $p(t)$ is given by the scheduling map:
	\begin{align}
		\begin{split}
		p(t) = \bar{f}(z(t)) = \bar{f}(C_\mathrm{z} x(t) \!+\! D_\mathrm{zu} \tilde{u}(t)).
		\end{split}
	\end{align}
	
	This results in the following affine state-space LPV structure:
	\begin{align} \label{eq:AffineLPV_LFR}
		\begin{split}
		\dot{x}(t) &= (A + A_\mathrm{p}(p(t))) x(t) + (B_\mathrm{u} + B_\mathrm{p}(p(t))) \tilde{u}(t), \\
		\tilde{y}(t)   &= (C_\mathrm{y} + C_\mathrm{p}(p(t))) x(t) + (D_\mathrm{yu} + D_\mathrm{p}(p(t))) \tilde{u}(t),
		\end{split}
	\end{align}
	with:
	\begin{align} \label{eq:LPVMatrices}
		\begin{split}
		A_\mathrm{p}(p(t)) = B_\mathrm{w}p(t) C_\mathrm{z}, \quad C_\mathrm{p}(p(t)) = D_\mathrm{yw}p(t) C_\mathrm{z},\\
		B_\mathrm{p}(p(t)) = B_\mathrm{w}p(t) D_\mathrm{zu}, \quad	D_\mathrm{p}(p(t)) = D_\mathrm{yw}p(t) D_\mathrm{zu}.\\
		\end{split}
	\end{align}
	Note that the expressions in Eq.~\eqref{eq:LPVMatrices} are linear in $p(t)$.
	
	\subsection{Remarks} \label{sec:remarks}
	The resulting affine state-space LPV representation is not unique, a state transformation can be introduced. Also the latent variables $z(t)$ and $w(t)$ are non-uniquely defined, similar to the non-uniqueness that is present in block-oriented systems \citep{SchoukensM2015a,SchoukensM2015c}. This non-uniqueness can be further explored for the scheduling variables selection. From a control design point of view it is desirable for the scheduling signals $p(t)$ or the input of the static nonlinearity $z(t)$ to be measurable or observable. This will be explored in future research.

	Note that the proposed algorithm, including the constant offset removal and the nonlinear function factorization can be completely automated starting from a NLFR representation. However, in some cases it can be worthwhile to search for an 'optimal' factorization. As discussed above, multiple factorizations of a multivariate nonlinear function are possible. Hence, the search for an 'optimal' factorization is more difficult to automate and requires more study.

\section{Simulation Example: 2-DOF Nonlinear Mass-Spring-Damper System} \label{sec:simulation}
	
	\subsection{Nonlinear System Equations}
\begin{figure}[bt]
			\centering
				\includegraphics[width=0.75\columnwidth]{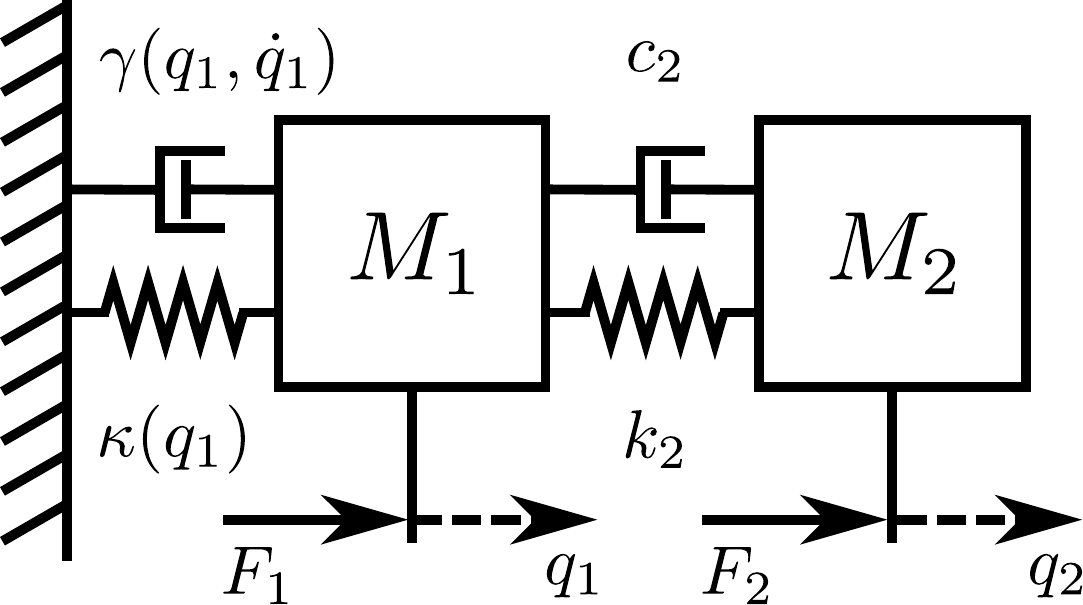}
			\caption{The considered 2 DOF mass-spring-damper system where $y_i$ denotes the position, and $F_i$ the force applied on mass $M_i$.}
			\label{fig:MSD}
		\end{figure}

		A 2-DOF mass-spring-damper system with unit masses ($M_1 = M_2 = 1$ in Figure~\ref{fig:MSD}) a nonlinear spring $\kappa(y_1)$ and nonlinear position-dependent damping $\gamma(y_1,\dot{y}_1)$ is considered:	
		\begin{align}
		\begin{split}
			\begin{bmatrix}
				\ddot{q}_1 \\ \ddot{q}_2
			\end{bmatrix}
			+
			\begin{bmatrix}
				c2 & -c2 \\
				-c2 & c2
			\end{bmatrix}
			\begin{bmatrix}
				\dot{q}_1 \\ \dot{q}_2
			\end{bmatrix}
			+
			\begin{bmatrix}
				k2 & -k2 \\
				-k2 & k2
			\end{bmatrix}
			\begin{bmatrix}
				q_1 \\ q_2
			\end{bmatrix}  \\
			+
			\begin{bmatrix}
				\gamma(q_1,\dot{q}_1) + \kappa(q_1) \\
				0
			\end{bmatrix}
			=
			\begin{bmatrix}
				F_1 \\ F_2
			\end{bmatrix}	
		\end{split}					
		\end{align}
		where the nonlinear damping and spring are given by:
		\begin{align}
		\begin{split}
			\gamma(q_1,\dot{q}_1) &= c_1 \dot{q}_1 \!+\! c_3 \sin(c_4 q_1) \dot{q}_1 \!+\! c_5 \dot{q}_1^2  \\
			\kappa(q_1) &= k_1 q_1 \!+\! k_3 q_1^3,
		\end{split}
		\end{align}
		with $k_1 = \pi^2$, $k_2 = (1.2\pi)^2$, $k_3 = 10$, $c_1 = 0.1$, $c_2 = 0.01$, $c_3 = 0.1$, $c_4 = 10$, $c_5 = 0.2$. 
	
	\subsection{NLFR Representation}
		
		A NLFR representation can be obtained using the notation as in Eq.~\eqref{eq:LFRSS1}:
		\begin{align}
			\begin{split}
			\dot{x} &= A x + \begin{bmatrix} B_\mathrm{w} & B_\mathrm{u} \end{bmatrix} \begin{bmatrix}w \\ u\end{bmatrix}, \\
			\begin{bmatrix}z \\ y\end{bmatrix} &= \begin{bmatrix} C_\mathrm{z} \\ C_\mathrm{y} \end{bmatrix}  x,
			\end{split}
		\end{align}
		where
		\begin{align}
			\begin{split}
			u &= \begin{bmatrix} F_1 \\ F_2 \end{bmatrix}, \quad y = \begin{bmatrix} q_1 \\ q_2 \end{bmatrix}, \\
			x & = \begin{bmatrix} q_1 & q_2 & \dot{q}_1 & \dot{q}_2 \end{bmatrix}^\top, \\
			z &= \begin{bmatrix} x_1 \\ x_3 \end{bmatrix}, \\
			w &= f(z) \\
				 &=c_3 \sin(c_4 z_1) z_2 + c_5 z_2^2 + k_3 z_1^3,
			\end{split}
		\end{align}
		and
		\begin{align}
			\begin{split}
			A &= \begin{bmatrix} 
					0 	    & 0  & 1 	   & 0 \\
					0 	    & 0  & 0 	   & 1 \\
					-k_1-k_2& k_2& -c_1-c_2& c_2 \\
					k_2 	&-k_2& c_2     &-c_2 \\
				\end{bmatrix}, \\
			B_\mathrm{w} &= \begin{bmatrix} 
					0 \\
					0 \\
					-1 \\
					0
				\end{bmatrix}, \quad
			B_\mathrm{u} = \begin{bmatrix} 
								0 & 0 \\
								0 & 0 \\
								1 & 0 \\
								0 & 1
							\end{bmatrix}, \\
			C_\mathrm{z} &= \begin{bmatrix}  
					1 & 0 & 0 & 0 \\
					0 & 0 & 1 & 0  
				\end{bmatrix}, \quad
			C_\mathrm{y} = \begin{bmatrix} 
					1 & 0 & 0 & 0 \\
					0 & 1 & 0 & 0  
				\end{bmatrix},			
			\end{split}
		\end{align}
		where $z_i$ denotes the $i$-th entry from the vector $z$ and the matrices $D_\mathrm{zw}$, $D_\mathrm{zu}$, $D_\mathrm{yw}$, $D_\mathrm{yu}$ are all zero matrices.
		
	\subsection{LPV Embedding} \label{sec:MSDEmbedding}
		It can be observed that no offset handling is required since it holds for the nonlinearity in the NLFR that $f(0) = 0$.
		
		Multiple factorizations are possible using the factorization scheme presented in this paper. By first factorizing $z_1$, and $z_2$ next, we obtain:
		\begin{align} \label{eq:schedulingmap}
		\begin{split}
			f(z) &= \bar{f}(z)z = \begin{bmatrix} \bar{f}_1(z_1) & \bar{f}_2(z_1,z_2) \end{bmatrix} \begin{bmatrix} z_1 \\ z_2 \end{bmatrix} \\
				 &= \begin{bmatrix} k_3 z_1^2 \\ c_3 \sin(c_4 z_1) + c_5 z_2 \end{bmatrix}^\top \begin{bmatrix} z_1 \\ z_2 \end{bmatrix}.
		\end{split}
		\end{align} 
		Note that the functions $k_3 z_1^2$ and $c_3 \sin(c_4 z_1) + c_5 z_2$ are the resulting scheduling maps of the LPV embedding.
		
		Another possibility (first factorize $z_2$, and $z_1$ next) is: 		
		\begin{align}
			f(z) &= \begin{bmatrix} k_3 z_1^2 + c_3 \sinc(c_4 z_1)z_2 \\ c_5 z_2 \end{bmatrix}^\top \begin{bmatrix} z_1 \\ z_2 \end{bmatrix}.
		\end{align}
		Many more factorizations are possible, beyond the ones obtained with the proposed factorization scheme. For instance:
		\begin{align}
			f(z) \!&=\! \begin{bmatrix} k_3 z_1^2 \!+\! (1\!-\!\alpha) c_3 \sinc(c_4 z_1)z_2 \\ \alpha c_3 \sin(c_4 z_1) \!+\! c_5 z_2 \end{bmatrix}^\top \!\!\! \begin{bmatrix} z_1 \\ z_2 \end{bmatrix}
		\end{align} 
		for any finite $\alpha$ results in an equivalent representation.
		
		Using the choice \eqref{eq:schedulingmap}, the equivalent LPV representation of the considered system is given by:
		\begin{align}
			\begin{split}
			\dot{x} &= A x + A_\mathrm{p}(p(t)) x + B_\mathrm{u} u, \\
			y &= C_\mathrm{y} x,
			\end{split}
		\end{align}
		where
		\begin{align}
			A_\mathrm{p}(p(t)) = \begin{bmatrix} 0& 0& 0& 0 \\ 0& 0& 0& 0\\ -p_1(t)& 0& -p_2(t)& 0\\ 0& 0& 0& 0 \end{bmatrix},
		\end{align}
		 with $p_1(t) = \bar{f}_1(q_1(t))$, $p_2(t) = \bar{f}_2(q_1(t),\dot{q}_1(t))$ provided in Eq.~\eqref{eq:schedulingmap}.
		
	\subsection{Results}
\begin{figure}[bt]
			\centering
				\includegraphics[width=1\columnwidth]{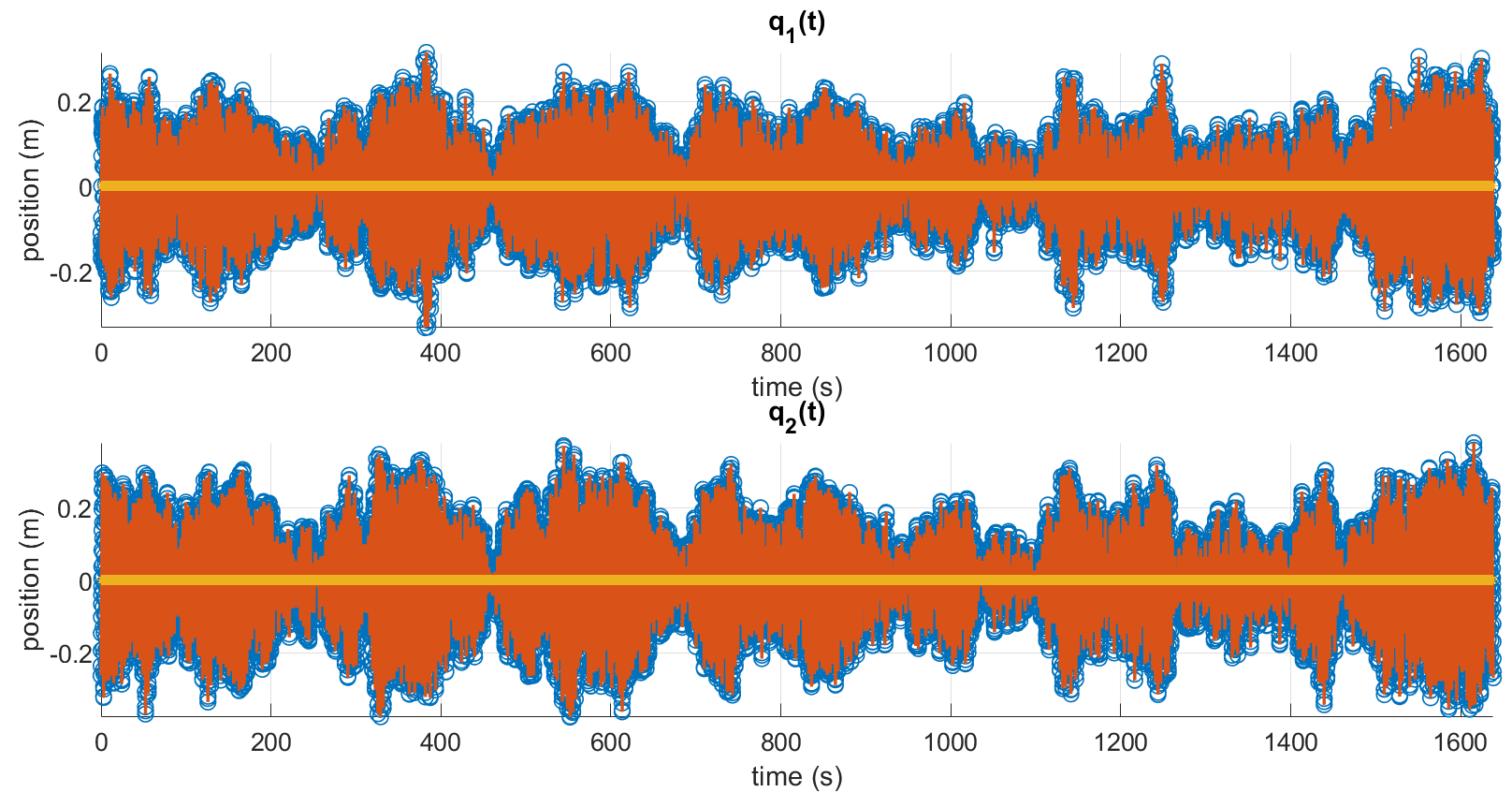}
			\caption{Time domain output: the NLFR output, $y_1 = q_1$ and $y_2 = q_2$, (blue circles) coincides perfectly with the embedded NLFR output (red line), as is illustrated by the residual error (orange) that is equal to zero or below Matlab precision.}
			\label{fig:SimEx_OutputTime}
		\end{figure}

\begin{figure}[bt]
			\centering
				\includegraphics[width=1\columnwidth]{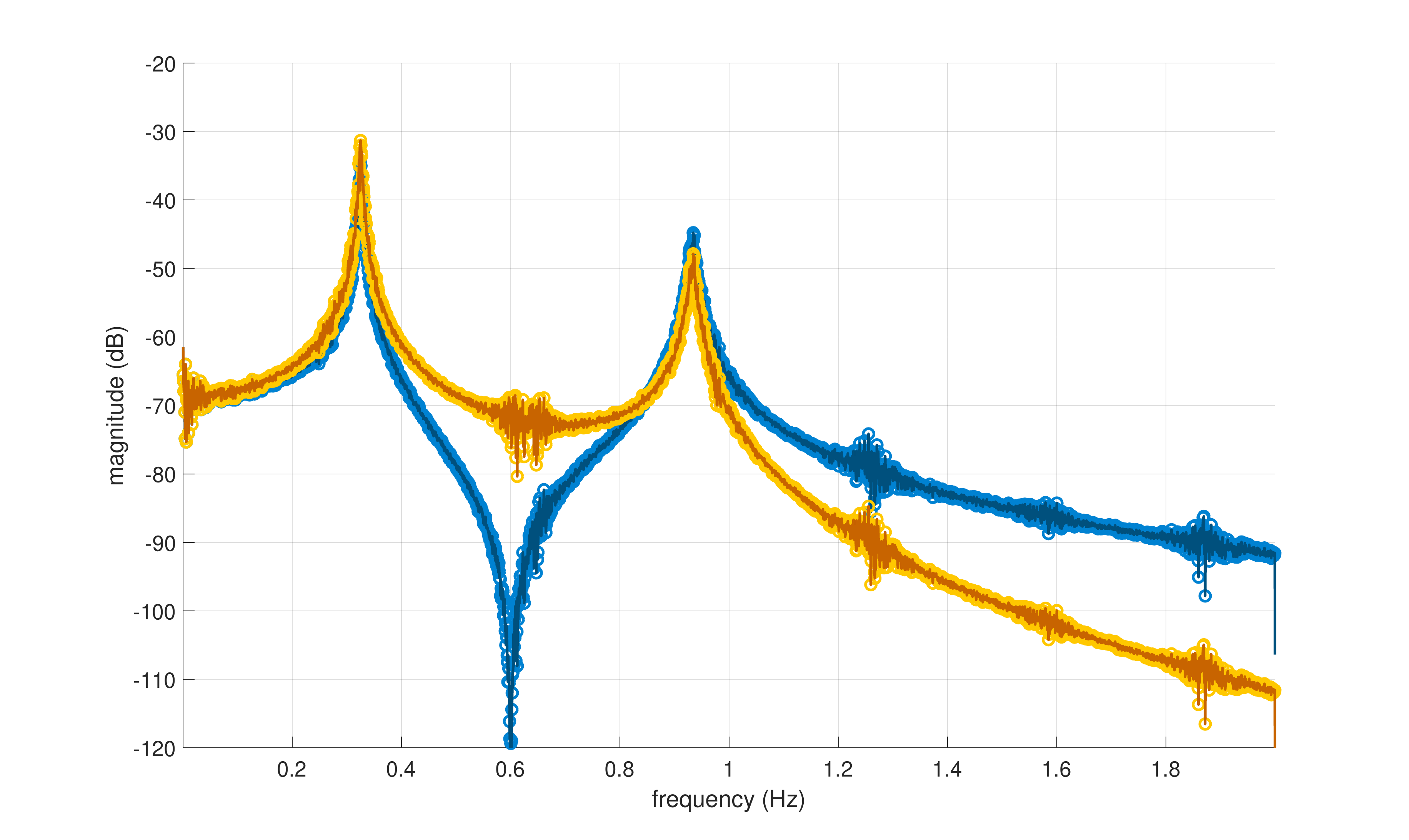}
			\caption{Frequency domain output ($y_1 = q_1$ in blue, and $y_2 = q_2$ in orange): the NLFR output (circles, lighter color) coincides perfectly with the embedded NLFR output (line, darker color)}
			\label{fig:SimEx_OutputFreq}
		\end{figure}

		The output responses of the NLFR representation and by the embedding obtained LPV representation are depicted in the time- and frequency domain in Figures~\ref{fig:SimEx_OutputTime} and \ref{fig:SimEx_OutputFreq} respectively. A perfect match between both can be observed.
		
		The scheduling signals $p$ are obtained by applying the scheduling map on the trajectories $x$ and $u$ obtained by the nonlinear simulation.
		
\begin{figure}[bt]
			\centering
				\includegraphics[width=0.95\columnwidth]{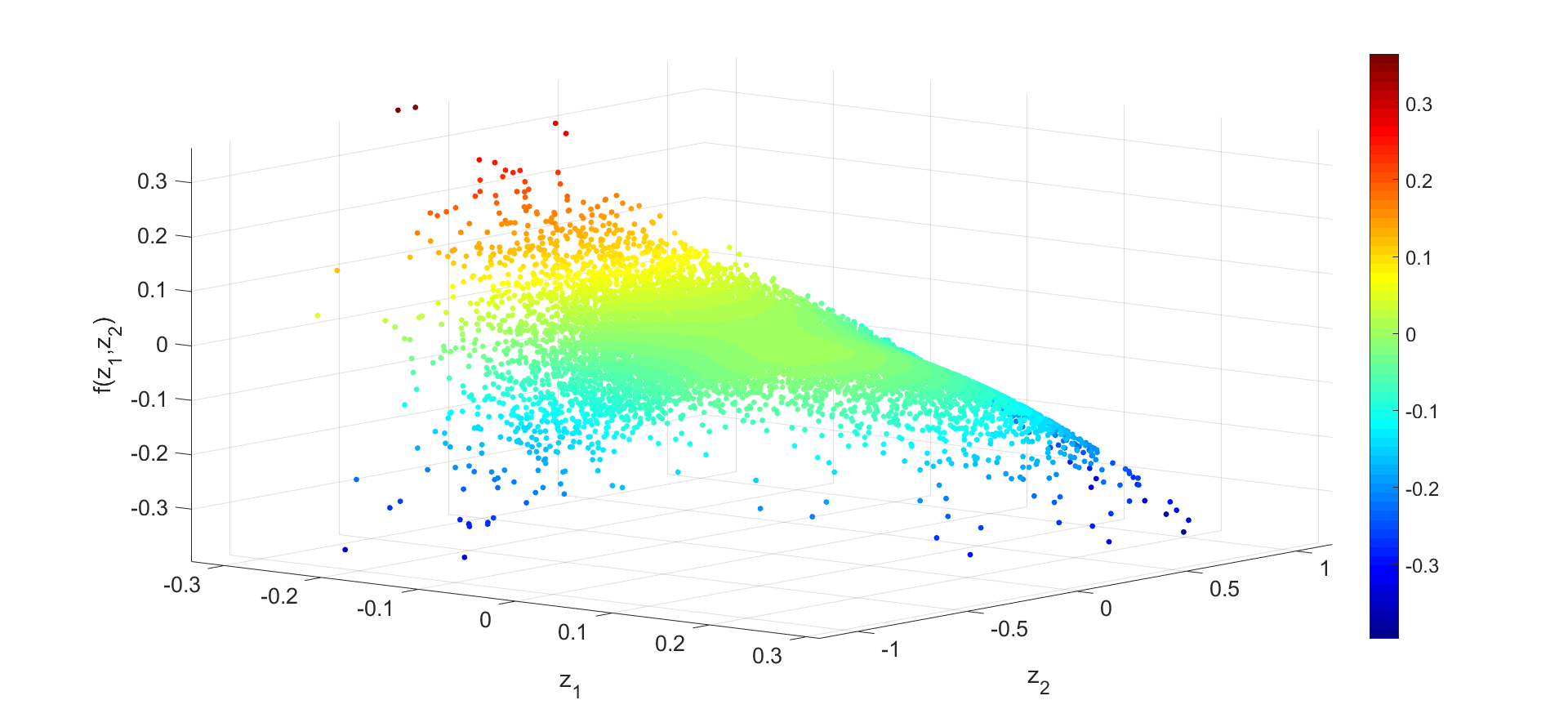}
			\caption{Surface plot of the original nonlinearity: $c_3 \sin(c_4 q_1) \dot{q}_1 + c_5 \dot{q}_1^2 + k_3 q_1^3$.}
			\label{fig:SimEx_NL}
		\end{figure}

\begin{figure}[bt]
			\centering
				\includegraphics[width=0.95\columnwidth]{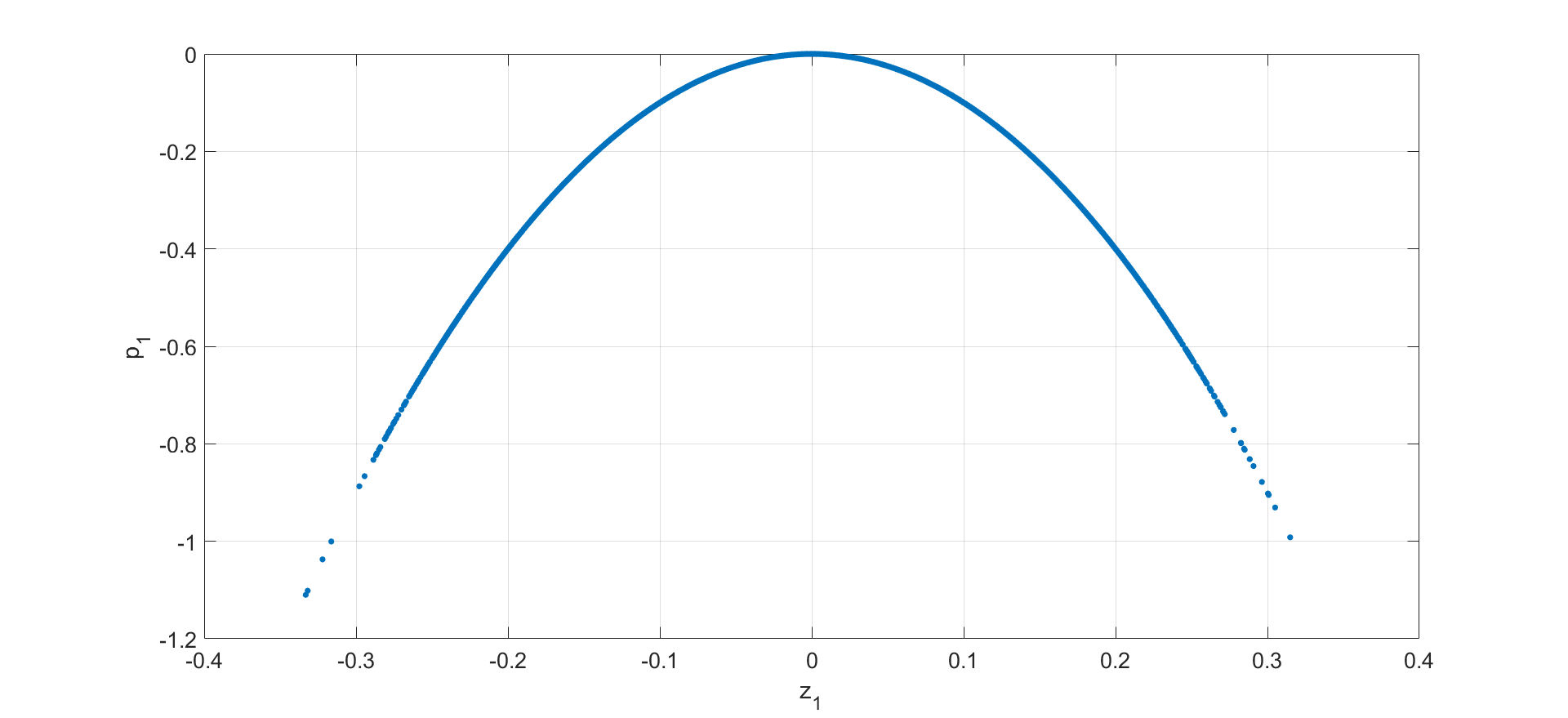}
			\caption{The scheduling map (factorized nonlinearity $\bar{f}_1(z_1)$) of $p_1$ in the LPV representation.}
			\label{fig:SimEx_f1}
		\end{figure}

\begin{figure}[bt]
			\centering
				\includegraphics[width=0.95\columnwidth]{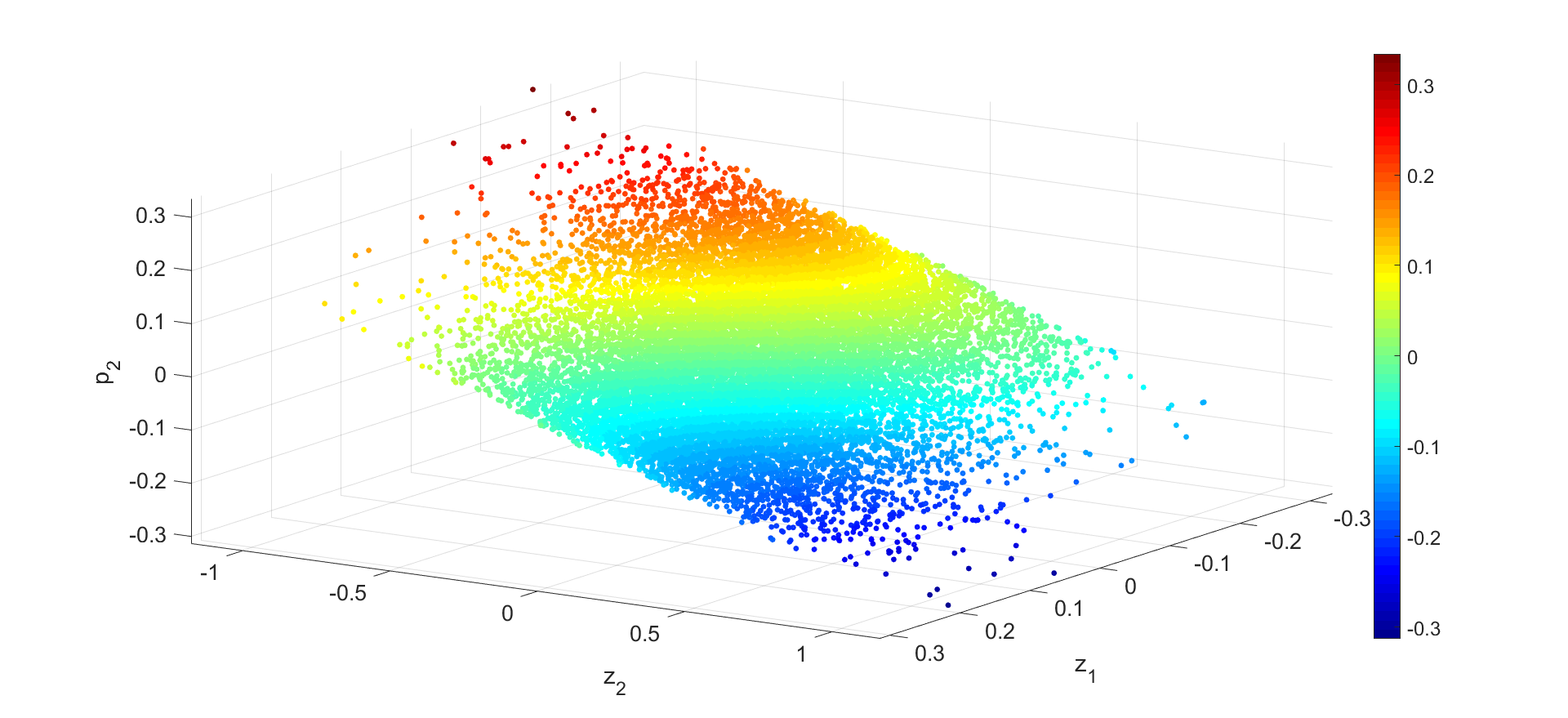}
			\caption{The scheduling map (factorized nonlinearity $\bar{f}_2(z_1,z_2)$) of $p_2$ in the LPV representation.}
			\label{fig:SimEx_f2}
		\end{figure}

		The system nonlinearity that is present in the NLFR representation (see Figure~\ref{fig:SimEx_NL}) is factorized in 2 nonlinear functions $\bar{f}_1(z_1)$, $\bar{f}_2(z_1,z_2)$ shown in Figures~\ref{fig:SimEx_f1} and \ref{fig:SimEx_f2} respectively.
		
\section{Conclusion} \label{sec:Conclusions}
	This paper demonstrates how the class of MIMO NLFR systems can be exactly represented by an affine state-space LPV model under 3 mild assumptions. A systematic and automated embedding procedure is proposed for the underlying conversion problem. By this procedure an NLFR system is embedded in an LPV representation without introducing any new singularities. The effectiveness of the proposed approach is illustrated on a 2-DOF nonlinear mass-spring-damper example.

	Further research will explore the optimal selection of the scheduling map based upon the various possibilities provided by the factorization, by taking into account the control objectives in the modeling process, in terms of achievable performance and measures of conservativeness.

\bibliography{ReferencesLibraryV2}                                

\end{document}